\documentclass{edm_article}

\usepackage{comment}
\usepackage{hyperref}
\usepackage{booktabs}
\usepackage{amsmath}
\newcommand\norm[1]{\left\lVert#1\right\rVert}
\usepackage{subcaption}
\usepackage{bbm}
\usepackage{pgfplots}

\begin{document}

\title{Knowledge Transfer by Discriminative Pre-training for Academic Performance Prediction}
%
\numberofauthors{1}
\author{
Byungsoo Kim, Hangyeol Yu, Dongmin Shin, Youngduck Choi\\
\affaddr{Riiid! AI Research}\\
\email{\{byungsoo.kim,hangyeol.yu,dm.shin,youngduck.choi\}@riiid.co}
}

\maketitle


\begin{abstract}
The needs for precisely estimating a student’s academic performance have been emphasized with an increasing amount of attention paid to Intelligent Tutoring System (ITS).
However, since labels for academic performance, such as test scores, are collected from outside of ITS, obtaining the labels is costly, leading to label-scarcity problem which brings challenge in taking machine learning approaches for academic performance prediction.
To this end, inspired by the recent advancement of pre-training method in natural language processing community, we propose DPA, a transfer learning framework with \textbf{D}iscriminative \textbf{P}re-training tasks for \textbf{A}cademic performance prediction.
DPA pre-trains two models, a generator and a discriminator, and fine-tunes the discriminator on academic performance prediction.
In DPA’s pre-training phase, a sequence of interactions where some tokens are masked is provided to the generator which is trained to reconstruct the original sequence.
Then, the discriminator takes an interaction sequence where the masked tokens are replaced by the generator’s outputs, and is trained to predict the originalities of all tokens in the sequence.
Compared to the previous state-of-the-art generative pre-training method, DPA is more sample efficient, leading to fast convergence to lower academic performance prediction error.
We conduct extensive experimental studies on a real-world dataset obtained from a multi-platform ITS application and show that DPA outperforms the previous state-of-the-art generative pre-training method with a reduction of 4.05\% in mean absolute error and more robust to increased label-scarcity.
\end{abstract}

%

\keywords{Academic Performance Prediction, Deep Learning, Transfer Learning, Discriminative Pre-training}

\section{Introduction}

\begin{figure}[h]
\centering
\includegraphics[width=0.5\textwidth]{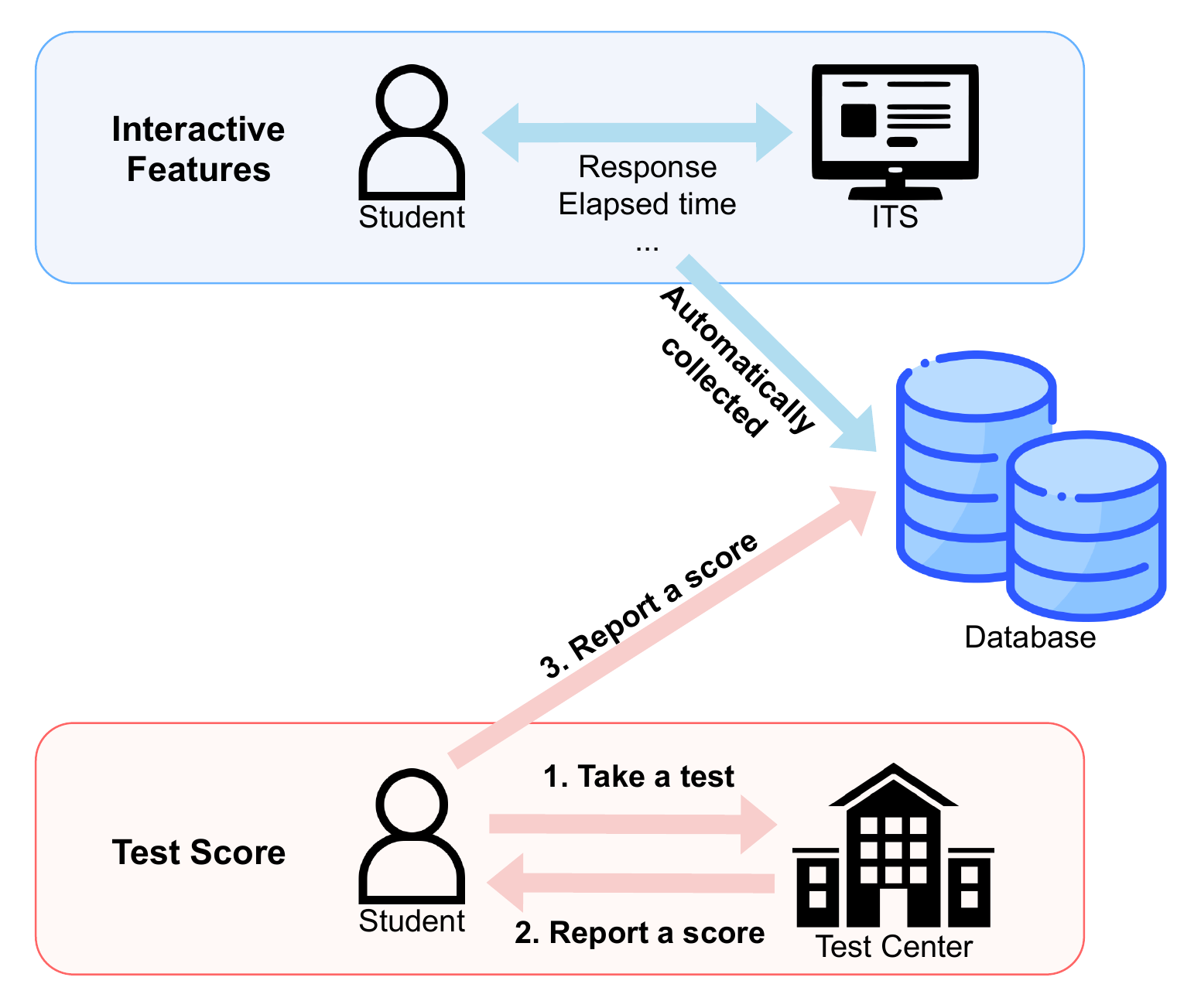}
\caption{Interactive features, such as student response and elapsed time for the response, are automatically recorded to the database whenever a student interacts with ITS.
On the other hand, more complicated steps are necessary to obtain a test score: a student should take the test in the designated test center, receive the test score, and report the score to ITS.}
\label{fig:label_scarce}
\end{figure}

Predicting a student’s future academic performance is a fundamental task for developing modern Intelligent Tutoring System (ITS) which aims to provide personalized learning experience by supporting educational needs of each individual.
However, labels for academic performance, such as test scores, are often scarce since they are external to ITS.
For example, as shown in Figure \ref{fig:label_scarce}, test scores are not automatically collected inside of ITS.
Obtaining a test score requires a student to take the test in the designated test center, receive the score, and report the score to ITS.
Transfer learning is a commonly taken approach to address such label-scarcity problems across different domains of machine learning.
In this framework, a model is first pre-trained to optimize auxiliary objectives with abundant data, and then fine-tuned on the task of interest.
In Artificial Intelligence in Education (AIEd) community, \cite{choi2020assessment} introduced Assessment Modeling (AM), a set of pre-training tasks for label-scarce educational problems including academic performance prediction.
AM proposed a pre-training method where first, a masked interaction sequence is generated by replacing a set of interactive features which can serve as criteria for pedagogical evaluation with artificial mask tokens.
Then, given the masked interaction sequence, a model is pre-trained to predict the masked interactive features.
The idea was borrowed from the Masked Language Modeling (MLM) pre-training method proposed in \cite{devlin2018bert}.
In the MLM pre-training method, given a masked word sequence where some words in the sequence are replaced with an artificial mask token, a model is pre-trained to predict the masked words.
However, recently, \cite{clark2020electra} pointed out that the MLM pre-training method has poor sample efficiency and suffers from pre-train/fine-tune discrepancy due to the artificial mask token, and proposed a new discriminative pre-training method.
Considering the problems are also inherent in AM, potential gains are expected to be obtainable when the discriminative pre-training method is applied to academic performance prediction.

To this end, we propose DPA, a transfer learning framework with \textbf{D}iscriminative \textbf{P}re-training tasks for \textbf{A}cademic performance prediction.
There are two models in DPA: a generator and a discriminator.
In DPA’s pre-training phase, the generator is trained to predict the masked interactive features in the same way as AM.
Then, given a replaced interaction sequence which is generated by replacing the masked features with the generator’s outputs, the discriminator is trained to predict whether each token in the sequence is the same as the one in the original interaction sequence.
After the pre-training, the generator is thrown away and only the discriminator is fine-tuned on academic performance prediction.
Also, we investigate diverse pre-training tasks for the generator and show that pre-training the generator to predict a student’s response is more effective than to predict the correctness and timeliness of their response which were considered as the most pedagogical interactive features in AM.
Extensive experimental studies conducted on a real-world dataset collected from a multi-platform ITS application show that DPA outperforms AM with a reduction of 4.05\% in mean absolute error and more robust when the degree of label-scarcity increases.
Through a series of ablation experiments, we show that DPA’s sample efficient pre-training contributed most to the improvement from AM to DPA.

\section{Santa: A Self-study Solution \\
Equipped with an AI Tutor for English Education}

In this paper, we conduct experiments on a real-world dataset obtained from \emph{Santa}\footnote{\url{https://aitutorsanta.com}}, a multi-platform ITS with more than a million users in South Korea available through Android,
iOS, and Web that exclusively focuses on the Test of English for International Communication (TOEIC) standardized examination.
The publicly accessible version of the dataset was released under the name \emph{EdNet} \cite{choi2020ednet}.
The TOEIC consists of two timed sections, Listening Comprehension (LC) and Reading Comprehension (RC).
There are a total of 100 multiple choice exercises in each section, and the total score for each section is 495 in steps of 5 points.
\emph{Santa} provides learning experiences of solving exercises, studying explanations, and watching lectures.
When a student consumes a specific learning content, \emph{Santa} diagnoses their current academic status based on their learning activities records and recommends another learning content appropriate for their current position.
\emph{Santa} records diverse types of interactive features, such as student response, the duration of time the student took to respond, and the time interval between the current and previous learning activities.
However, unlike the interactive features automatically collected from \emph{Santa}, obtaining the official TOEIC score requires more steps: a student should register and pay for the test, take the test in the designated test center, receive the test score from the Educational Testing Service, and report the score to \emph{Santa} (Figure \ref{fig:label_scarce}).
\emph{Santa} collected students’ TOEIC score data by offering small gifts to students when they report their scores.

\section{Transfer Learning for Academic Test Performance Prediction}
\label{sec:transfer_learning}

\begin{figure*}[t]
\centering
\includegraphics[width=1\textwidth]{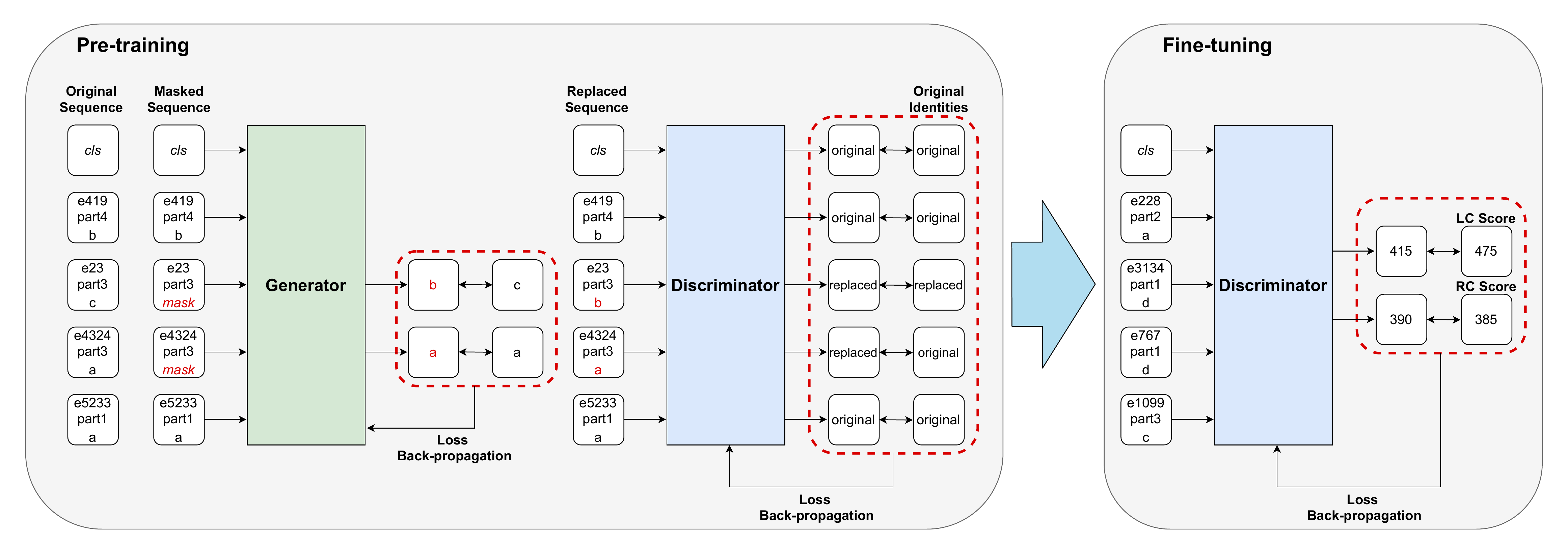}
\caption{The overall pre-training/fine-tuning process of DPA when each token in an interaction sequence is a set of \emph{eid}, \emph{part}, and \emph{response}, and \emph{response} is a feature being masked.
\emph{mask} and \emph{cls} are special tokens for mask and classification, respectively, which are the same as the ones used in \cite{devlin2018bert}.}
\label{fig:overall_architecture}
\end{figure*}

To overcome the label-scarcity problem in academic test performance prediction, we consider burgeoning machine learning discipline of transfer learning.
There is an open issue of what information to transfer or which pre-training task is the most effective for academic test performance prediction.
Previous studies proposed two types of pre-training methods for AIEd Tasks: interaction-based method which models students’ dynamic learning behaviors \cite{guo2015predicting,hunt2017transfer,ding2019transfer,choi2020assessment}, and content-based method which learns representations of learning contents \cite{huang2017question,su2018exercise,liu2019ekt,sung2019improving,yin2019quesnet}.
\cite{choi2020assessment} showed that interaction-based pre-training method outperforms content-based pre-training methods when the pre-trained model is fine-tuned on several label-scarce educational tasks including academic test performance prediction.
Following this line of research, we propose a transfer learning framework where a model is pre-trained using only student interaction data, and fine-tune the pre-trained model on academic test performance prediction.
In this paper, we consider the following interactive features:
\begin{itemize}
\item \emph{eid}: A unique ID assigned to an exercise solved by a student.
There are a total of 14419 exercises in the dataset.
\item \emph{part}: Each exercise belongs to a specific part that represents the type of the exercise.
There are a total of 7 parts in the TOEIC.
\item \emph{response}: Since the TOEIC consists of multiple choice exercises and there are four options for each exercise, a student response for a given exercise is one of the options, ‘a’, ‘b’, ‘c’, or ‘d’.
\item \emph{correctness}: Whether a student responded correctly to a given exercise.
Note that \emph{correctness} is a coarse version of \emph{response} since \emph{correctness} is processed by comparing \emph{response} with a correct answer for a given exercise.
\item \emph{elapsed\_time}: The amount of time a student spent on solving a given exercise.
\item \emph{timeliness}: Whether a student responded to a given exercise under the time limit.
Note that \emph{timeliness} is a coarse version of \emph{elapsed\_time} since \emph{timeliness} is processed by comparing \emph{elapsed\_time} with the time limit recommended by domain experts for a given exercise.
\item \emph{exp\_time}: The amount of time a student spent on studying an explanation for an exercise they had solved.
\item \emph{inactive\_time}: The time interval between the current and previous interactions.
\end{itemize}
In our experiments, we normalize the values of \emph{elapsed\_time}, \emph{exp\_time}, and \emph{inactive\_time} so they are between 0 and 1 to stabilize the training process.

\section{Proposed Method}

Figure \ref{fig:overall_architecture} depicts our proposed method.
There are two models in DPA: a generator and a discriminator.
In pre-training phase, given a sequence of interactions $I = [I_1, \dots, I_T]$, where each interaction $I_t = \{f^1_t, \dots, f^k_t\}$ is a set of interactive features $f^i_t$, such as \emph{eid}, \emph{part}, and \emph{response}, a masked interaction sequence $I^M = [I^M_1, \dots, I^M_T]$ is generated by first randomly selecting a set of positions to mask $M = \{M_1, \dots, M_m\}$ $(m < T)$, and for the masked position $M_i$, masking out a fixed set of features $\{f^1_{M_i}, \dots, f^n_{M_i}\}$ $(n < k)$.
For instance, in Figure \ref{fig:overall_architecture}, if the original interaction sequence is [(e419, part4, b), (e23, part3, c), (e4324, part3, a), (e5233, part1, a)] where each token in the sequence is a set of \emph{eid}, \emph{part}, and \emph{response}, a masked interaction sequence where $M = \{2, 3\}$ and \emph{response} as a masked feature is [(e419, part4, b), (e23, part3, \emph{mask}), (e4324, part3, \emph{mask}), (e5233, part1, a)].
Then, the generator takes the masked interaction sequence $I^M$ as an input, and outputs predicted values $O^G_{ij}$ for the masked features $f^j_{M_i}$.
After that, a replaced interaction sequence $I^R = [I^R_1, \dots, I^R_T]$ is generated by replacing the masked features $f^j_{M_i}$ with the generator's predictions $O^G_{ij}$.
In Figure \ref{fig:overall_architecture}, since the generator's outputs for the masked features are ‘b’ and ‘a’, a replaced interaction sequence is [(e419, part4, b), (e23, part3, b), (e4324, part3, a), (e5233, part1, a)].
Then, the discriminator takes the replaced interaction sequence $I^R$ as an input, and predicts whether each token in the sequence is the same as the one in the original interaction sequence (original) or not (replaced).
After the pre-training, we throw away the generator and fine-tune the pre-trained discriminator on academic test performance prediction.
We provide detailed explanations of each component in the generator and the discriminator, and training objective functions in the following subsections.

\subsection{Interaction Embeddings}

The embedding layer produces a sequence of interaction embedding vectors by mapping each interactive feature to an appropriate embedding vector.
We take two different approaches to embed the interactive features depending on whether they are categorical (\emph{eid}, \emph{part}, \emph{response}, \emph{correctness}, and \emph{timeliness}) or continuous (\emph{elapsed\_time}, \emph{exp\_time}, and \emph{inactive\_time}) variables.
If an interactive feature is a categorical variable, we assign unique latent vectors to possible values of the feature including special values for mask (\emph{mask}) and classification (\emph{cls}).
Take \emph{response} as an example, there is an embedding matrix $E_{response} \in \mathbb{R}^{6 \times d_{emb}}$ where each row vector is assigned to one of ‘a’, ‘b’, ‘c’, ‘d’, \emph{mask}, and \emph{cls}.
If an interactive feature is a continuous variable, we assign a single latent vector to the feature.
Then, an embedding vector for the feature is computed by multiplying the latent vector and a value of the feature.
For instance, we compute an embedding vector for \emph{elapsed\_time} as $et * E_{elapsed\_time}$, where $et$ is a specific value and $E_{elapsed\_time} \in \mathbb{R}^{d_{emb}}$ is a latent vector assigned to \emph{elapsed\_time}.
Also, mask and classification for the continuous interactive features are indicated by setting their values to -1 and 0, respectively.
Not only embeddings for interactive features, positional embeddings are also incorporated into Transformer-based models \cite{vaswani2017attention} to consider chronological order of each token.
Rather than using conventional positional embeddings which stores an embedding vector for every possible position, we adopt axial positional embeddings \cite{kitaev2020reformer} to further reduce memory usage.
The final interaction embedding vector of dimension $d_{emb}$ for each time-step is the sum of all embedding vectors in the time-step.
The interaction embedding layer is shared by both the generator and the discriminator.

\subsection{Performer Encoder}

\begin{figure}[t]
\centering
\includegraphics[width=0.4\textwidth]{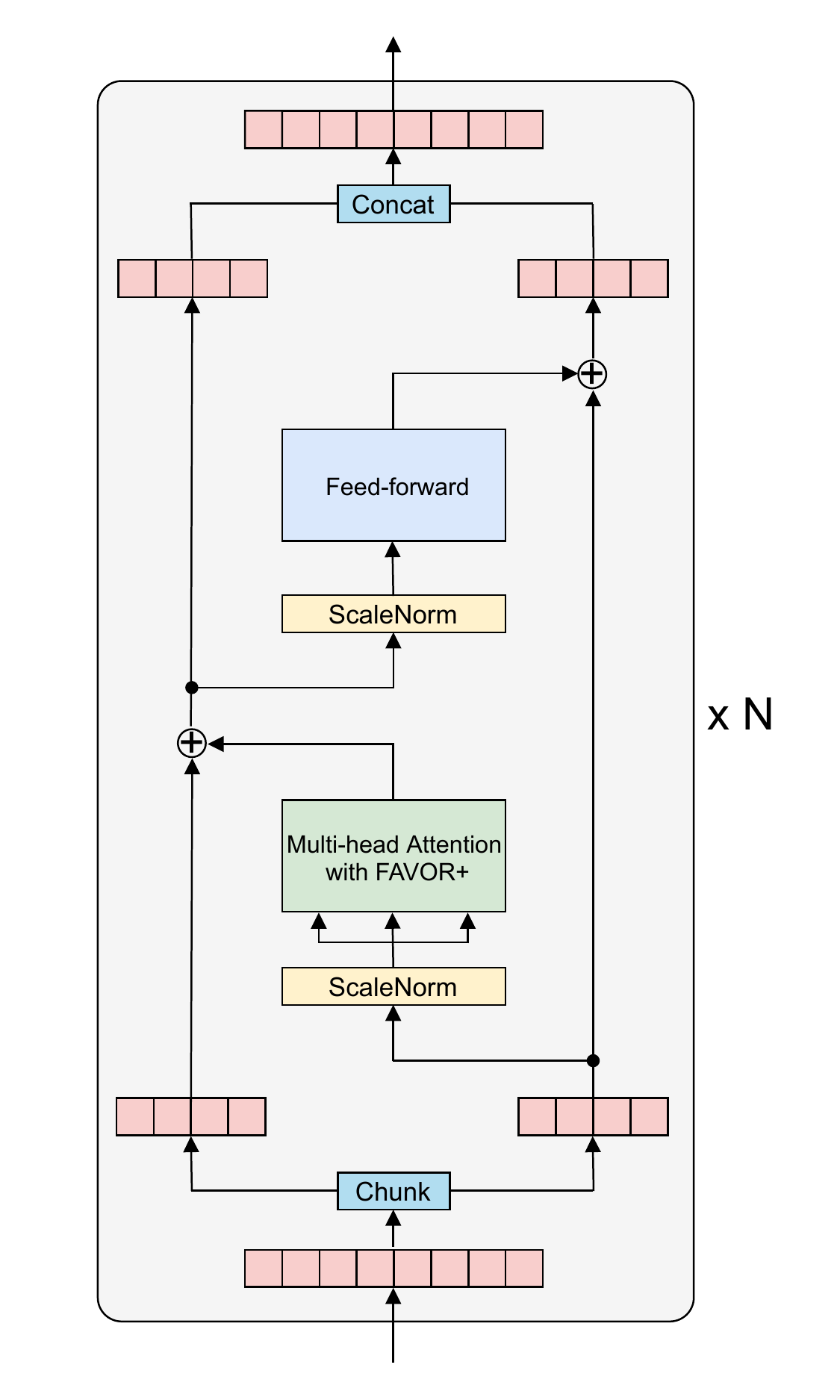}
\caption{The reversible layer in the Performer encoder is composed of the FAVOR+-based multi-head attention layer and the point-wise feed-forward layer.}
\label{fig:performer_encoder}
\end{figure}

\begin{figure*}[ht]
\centering
\includegraphics[width=1\textwidth]{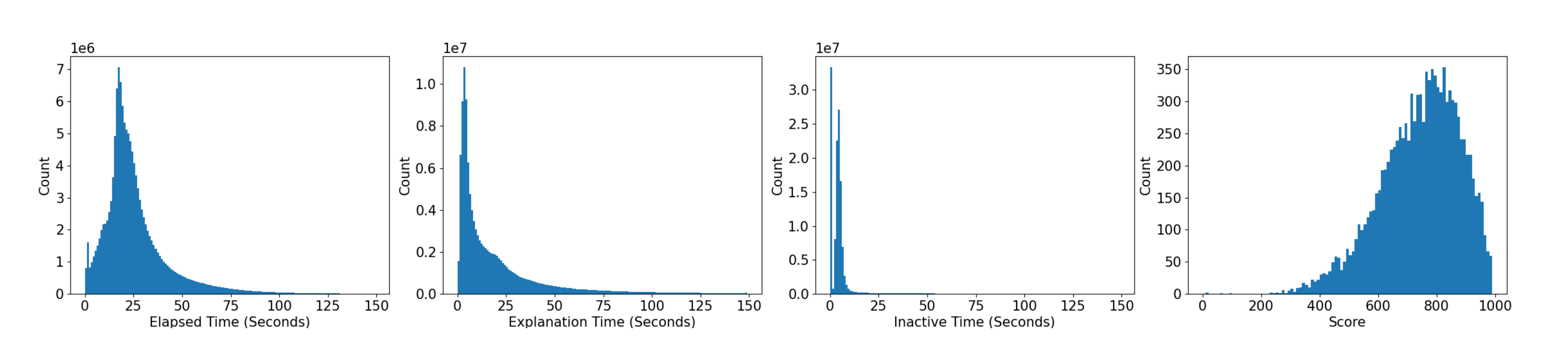}
\caption{Distributions of \emph{elapsed\_time}, \emph{exp\_time}, \emph{inactive\_time}, and score labels.}
\label{fig:data_distribution}
\end{figure*}

Since its successful debut in Natural Language Processing (NLP) community, Transformer’s attention mechanism has become a common recipe adopted across different domains of machine learning including speech processing \cite{li2019neural}, computer vision \cite{chen2020generative,dosovitskiy2020image}, and AIEd \cite{pandey2019self,choi2020towards,ghosh2020context,shin2020saint+}.
Compared to Recurrent Neural Network (RNN) family models, Transformer’s attention mechanism has benefits of capturing longer-range dependencies and allowing parallel training, which enables the model to achieve better performance with less training time.
However, despite these advantages, the time and memory complexities of computing the attention grow quadratically with respect to input sequence length, requiring demanding computing resources for training the model on long sequences.
For instance, if $L$ is input sequence length and $d$ is dimension of query, key, and value vectors, Transformer’s attention is computed as follows:
\begin{align*}
    \text{Attention}(Q, K, V) = \text{softmax}\Big(\frac{QK^\top}{\sqrt{d}}\Big)V,
\end{align*}
where $Q, K, V \in \mathbb{R}^{L \times d}$.
The time and memory complexities for computing $QK^\top$ in the above equation are $O(L^2d)$ and $O(L^2)$, respectively.
Therefore, the cost for training Transformer becomes prohibitive with large $L$, preventing training the model even on a single GPU.

The problem of improving the efficiency of Transformer's attention mechanism is a common concern of machine learning community.
Recent studies have proposed several methods to reduce the computing complexities lower than the quadratic degree with respect to input sequence length \cite{kitaev2020reformer,wang2020linformer,katharopoulos2020transformers,tay2020sparse,choromanski2020rethinking}.
In this paper, we adopt Performer \cite{choromanski2020rethinking} since it uses reasonable memory and makes a better trade-off between speed and performance \cite{tay2020long}.
Performer approximates attention kernels through Fast Attention Via positive Orthogonal Random features (FAVOR+) approach.
The R+-part in FAVOR+ computes $Q’$ and $K’$ by applying random feature map $\phi: \mathbb{R}^d \rightarrow \mathbb{R}^r_+$ to each query vector $q$ and key vector $k$ in $Q$ and $K$, respectively.
\begin{align*}
    Q' = [\phi(q_i), \dots, \phi(q_L)], K' = [\phi(k_i), \dots, \phi(k_L)] \in \mathbb{R}^{L \times r}.
\end{align*}
The FA-part leads to efficient attention mechanism computed as follows:
\begin{align*}
    \text{FAVOR+}(Q, K, V) &= D^{-1}(Q'((K')^\top V)) \\
    D &= diag(Q'((K')^\top 1_L)),
\end{align*}
where $1_L \in \mathbb{R}^L$ is an all-ones vector of length $L$ and $diag$ is a diagonal
matrix with the input as the diagonal.
Unlike Transformer’s attention mechanism which computes $QK^\top$, applies the softmax, and multiplies the result and $V$, FAVOR+ first computes $K’^\top V$, and multiplies the result and $Q’$, reducing the quadratic time and memory complexities to linear ones with respect to input sequence length.
Also, FAVOR+ can model most attention kernels used in practice.
If $\phi$ is a function given as below, FAVOR+ approximates the softmax attention kernel used in Transformer's attention mechanism.
\begin{align*}
    \phi(x) = \mathbb{E}_{w \sim \mathbb{N}(0,I_d)}\left[\exp{\big(w^\top x - \frac{\norm{x}^2}{2}\big)}\right].
\end{align*}
Furthermore, the authors proposed a generalized attention kernel, an attempt to model a kernelizable attention mechanism beyond the commonly used softmax attention kernel, by defining $\phi$ as below.
\begin{align*}
    \phi(x) = \mathbb{E}_{w \sim \mathbb{N}(0,I_d)}\left[\text{ReLU}(w^\top x)\right],
\end{align*}
where ReLU is the widely used Rectified Linear Unit function.
In this paper, we use the generalized attention kernel instead of the softmax attention kernel since the former showed better performance in \cite{choromanski2020rethinking} and our preliminary experiments. 
Lastly, the O-part further reduces the variance of the estimator by making different random samples $w$s to be orthogonal to each other through the standard Gram-Schmidt orthogonalization process.
For those who want to know more about theoretical details of FAVOR+, please refer \cite{choromanski2020rethinking}.

With the efficient attention mechanism by FAVOR+, we propose the Performer encoder which is stacks of several identical reversible layers described in Figure \ref{fig:performer_encoder}.
The reversible layer is based on Reversible Transformer \cite{gomez2017reversible,kitaev2020reformer} architecture to further improve memory efficiency in back-propagation.
An input of the reversible layer $x \in \mathbb{R}^{L \times d_{hidden}}$ is first chunked to $x_1, x_2 \in \mathbb{R}^{L \times d_{hidden}/2}$.
Then, scaled $l_2$ normalization (ScaleNorm) \cite{nguyen2019transformers} and FAVOR+-based multi-head attention layer (MultiHeadAttn) are applied to $x_2$, and the result is added to $x_1$ to compute $y_1 \in \mathbb{R}^{L \times d_{hidden}/2}$.
\begin{align*}
    y_1 = x_1 + \text{MultiHeadAttn}(\text{ScaleNorm}(x_2)).
\end{align*}
After that, the scaled $l_2$ normalization and point-wise feed-forward layer (FeedForward) are applied to $y_1$, and the result is added to $x_2$, computing $y_2 \in \mathbb{R}^{L \times d_{hidden}/2}$.
\begin{align*}
    y_2 = x_2 + \text{FeedForward}(\text{ScaleNorm}(y_1)).
\end{align*}
An output of the reversible layer $y \in \mathbb{R}^{L \times d_{hidden}}$ is a concatenation of $y_1$ and $y_2$.
We stack the reversible layer multiple times to allow the final model to sufficiently represent underlying data distribution.

\subsection{Generator}

The generator computes hidden representations $[h^G_1, \dots, h^G_T]$ by feeding the masked interaction sequence $I^M$ to a series of the interaction embedding layer (InterEmbedding), a point-wise feed-forward layer (GenFeedForward1), the Performer encoder (GenPerformerEncoder), and another point-wise feed-forward layer (GenFeedForward2):
\begin{align*}
    & [I^{ME}_1, \dots, I^{ME}_T] = \text{InterEmbedding}([I^M_1, \dots, I^M_T]) \\
    & [h^{GF}_1, \dots, h^{GF}_T] = \text{GenFeedForward1}([I^{ME}_1, \dots, I^{ME}_T]) \\
    & [h^{GP}_1, \dots, h^{GP}_T] = \text{GenPerformerEncoder}([h^{GF}_1, \dots, h^{GF}_T]) \\
    & [h^G_1, \dots, h^G_T] = \text{GenFeedForward2}([h^{GP}_1, \dots, h^{GP}_T]),
\end{align*}
where $I^{ME}_t, h^G_t \in \mathbb{R}^{d_{emb}}$ and $h^{GF}_t, h^{GP}_t \in \mathbb{R}^{d_{gen\_hidden}}$.
Then, depending on whether the masked features are categorical or continuous variables, generator outputs are computed differently.
If the masked features are categorical variables, the outputs are sampled from a probability distribution defined by the following softmax layer:
\begin{align*}
    O^G_{ij} \sim P_G(f^j_{M_i} | I^M) = \text{softmax}(E_j h^G_{M_i}).
\end{align*}
If the masked features are continuous variables, the outputs are computed by the following sigmoid layer:
\begin{align*}
    O^G_{ij} = \text{sigmoid}(E_j^\top h^G_{M_i}).
\end{align*}
Similar to the case of categorical masked features, one can sample the outputs from a probability distribution defined by $I^M$ and parameters of the generator when the masked features are continuous variables.
For instance, the outputs can be sampled from the Gaussian distribution where the mean and the variance are determined by $I^M$ and the generator's parameters.
However, we make the outputs deterministic because sampling the outputs underperforms in our preliminary experiments when the masked features are continuous variables.

\subsection{Discriminator}
In pre-training, outputs of the discriminator $O^D = [O^D_1, \dots,$ $O^D_T]$ is computed by applying a series of the interaction embedding layer (InterEmbedding), a point-wise feed-forward layer (DisFeedForward1), the Performer encoder (DisPerformerEncoder), and another point-wise feed-forward layer (DisFeedForward2) to the replaced interaction sequence $I^R$:
\begin{align*}
    & [I^{RE}_1, \dots, I^{RE}_T] = \text{InterEmbedding}([I^R_1, \dots, I^R_T]) \\
    & [h^{DF}_1, \dots, h^{DF}_T] = \text{DisFeedForward1}([I^{RE}_1, \dots, I^{RE}_T]) \\
    & [h^{DP}_1, \dots, h^{DP}_T] = \text{DisPerformerEncoder}([h^{DF}_1, \dots, h^{DF}_T]) \\
    & [O^D_1, \dots, O^D_T] = \text{DisFeedForward2}([h^{DP}_1, \dots, h^{DP}_T]),
\end{align*}
where $I^{RE}_t \in \mathbb{R}^{d_{emb}}$, $h^{DF}_t, h^{DP}_t \in \mathbb{R}^{d_{dis\_hidden}}$, $O^{D}_t \in \mathbb{R}$, and the sigmoid is applied to the last layer of the discriminator.
After the pre-training, we slightly modify the discriminator by replacing the last layer with a layer having appropriate dimension for academic test performance prediction.

\subsection{Training Objectives}
The objective for pre-training is to minimize the following loss function:
\begin{align*}
    \sum_{i=1}^m \sum_{j=1}^n \text{GenLoss}(O^G_{ij}, f^j_{M_i}) + \lambda \sum_{t=1}^T \text{DisLoss}(O^D_t, \mathbbm{1}(I^R_t = I_t)),
\end{align*}
where GenLoss is the cross entropy (or mean squared error) loss function if the masked features are categorical (or continuous) variables, DisLoss is the binary cross entropy loss function, and $\mathbbm{1}$ is the identity function.
For ease of notation, we omit an index for each input sample in the above equation.
If there are more than one masked features in each time-step $(n > 1)$, the generator is trained under the multi-task leaning scheme.
The objective for fine-tuning is to minimize the mean squared error loss between the model's predictions and score labels.

\section{Experiments}

\subsection{Dataset}

The pre-training dataset consists of student interaction logs.
The statistics of the dataset are summarized in Table \ref{tab:statistics_pre-training}.
We exclude student interaction logs less than 15 in length to reduce noisy interactions from students using \emph{Santa} just for trying out.
Since most of the values of \emph{elapsed\_time}, \emph{exp\_time}, and \emph{inactive\_time} are distributed in head areas as shown in Figure \ref{fig:data_distribution}, we set their maximum to 300, 300, and 86400 seconds, respectively, and any values more than that are capped off to the maximum.
We further normalize the values so that they are between 0 and 1 by dividing them by the maximum.

\begin{table}[ht]
\caption{Statistics of pre-training dataset.}
\centering
\begin{tabular}{l|l}
\toprule
Statistics & Value \\
\toprule
Number of students & 436847 \\
Number of interactions & 135884952 \\
Minimum length of interactions & 15 \\
Maximum length of interactions & 76379 \\
Mean length of interactions & 311.06 \\
Median length of interactions & 50 \\
Correct response ratio & 0.66 \\
Timely response ratio & 0.72 \\
\bottomrule
\end{tabular}
\label{tab:statistics_pre-training}
\end{table}

\begin{table}[ht]
\caption{Statistics of fine-tuning dataset.}
\centering
\begin{tabular}{l|l}
\toprule
Statistics & Value \\
\toprule
Number of students & 6814 \\
Number of score labels & 11212 \\
Minimum length of interactions before the test & 10 \\
Maximum length of interactions before the test & 39040 \\
Mean length of interactions before the test & 1408.31 \\
Median length of interactions before the test & 736 \\
\bottomrule
\end{tabular}
\label{tab:statistics_fine-tuning}
\end{table}

The fine-tuning dataset consists of test scores and student interaction logs before the test.
Table \ref{tab:statistics_fine-tuning} summarizes the statistics of the dataset.
The number of score labels is far less than that of interactions in the pre-training dataset, which leads to the label-scarcity problem in academic test performance prediction.
The minimum length of interactions before the test is 10 because \emph{Santa} required students to solve at least 10 exercises before taking the test.
The mean and median length of interactions before the test are longer than those in the pre-training dataset since students who decide to report their scores tend to be more serious about studying with \emph{Santa}.
As shown in Figure \ref{fig:data_distribution}, most of the scores are distributed over the range from 700 to 900, and there are very few scores in the below 200 area.
This has to do with the distribution of students using \emph{Santa} whose initial and goal scores are usually in the range of 600 to 700 and higher than 800, respectively.
Since the score labels are few in number, we perform 5-fold cross-validation by dividing the fine-tuning dataset into 5 splits and using 3/5, 1/5, and 1/5 of the dataset for training, validation, and test, respectively.

\subsection{Training Details and Hyperparameters}

We use the Mean Absolute Error (MAE) as the metric for academic test performance prediction.
The list of hyperparameters and their values are described in Table \ref{tab:hyperparameters}.
For each pre-training evaluation, the pre-trained model is fine-tuned and cross-validated on the validation set, which results in the same number of validation results as the number of pre-training evaluations.
Then, we select the model with the best validation result and report an evaluation result of the model on the test set.

\begin{table}[ht!]
\caption{Pre-train/fine-tune hyperparameters.}
\centering
\begin{tabular}{l|l}
\toprule
Hyperparameter & Value \\
\toprule
Attention window size & 1024 \\
Masked interaction ratio & 0.6 \\
$\lambda$ & 1 \\
\midrule
\multicolumn{1}{l|}{\textbf{Embedding}} \\
Interaction embedding dimension & 256 \\
Axial positional embedding shape & [32, 32] \\
Axial positional embedding dimension & [64, 192] \\
\midrule
\multicolumn{1}{l|}{\textbf{Generator}} \\
Number of reversible layers & 4 \\
Hidden layer dimension & 64 \\
Hidden activation function & GELU \cite{hendrycks2016gaussian} \\
Hidden layer dropout probability & 0.1 \\
Number of attention heads & 2 \\
Attention head dimension & 64 \\
Attention dropout probability & 0.1 \\
Feed-forward intermediate layer dimension & 256 \\
\midrule
\multicolumn{1}{l|}{\textbf{Discriminator}} \\
Number of reversible layers & 4 \\
Hidden layer dimension & 256 \\
Hidden activation function & GELU \\
Hidden layer dropout probability & 0.1 \\
Number of attention heads & 8 \\
Attention head dimension & 64 \\
Attention dropout probability & 0.1 \\
Feed-forward intermediate layer dimension & 1024 \\
\midrule
\multicolumn{1}{l|}{\textbf{FAVOR+}} \\
Number of random features & 256 \\
Random features redrawing interval & 1000 \\
\midrule
\multicolumn{1}{l|}{\textbf{Optimization}} \\
Optimizer & Adam \cite{kingma2014adam} \\
Adam $\beta_1$ & 0.9 \\
Adam $\beta_2$ & 0.98 \\
Adam $\epsilon$ & 1e-09 \\
Scheduler & Noam \cite{vaswani2017attention} \\
Noam warm-up steps & 4000 \\
\midrule
\multicolumn{1}{l|}{\textbf{Pre-training}} \\
Batch size & 64 \\
Batch update steps before each evaluation & 5000 \\
Number of evaluations & 40 \\
\midrule
\multicolumn{1}{l|}{\textbf{Fine-tuning}} \\
Batch size & 64 \\
Batch update steps before each evaluation & 10 \\
Patience & 30 \\
\bottomrule
\end{tabular}
\label{tab:hyperparameters}
\end{table}

\subsection{Effects of Generator's Pre-training Tasks}

There are multiple interactive features to be masked in each token of the interaction sequence, which raises a question of how to construct a set of masked interactive features, and accordingly, which pre-training task for the generator is the most effective for academic test performance prediction.
By default, all interactive features listed in Section \ref{sec:transfer_learning} are taken as inputs for both the generator and discriminator.
However, if \emph{response} (or \emph{elapsed\_time}) is masked, \emph{correctness} (or \emph{timeliness}) is excluded from the inputs and vice versa since there is an overlap of information that the features represent.
For example, when both \emph{response} and \emph{correctness} are taken as inputs, and \emph{correctness} is masked, the generator can predict the masked \emph{correctness} by only looking at \emph{eid} and \emph{response} without considering other interactions, which leads to poor pre-training.
The results are described in Table \ref{tab:comparison_pre-training_tasks}.

\begin{table}[h]
\centering
\caption{Comparison between different pre-training tasks.}
\begin{tabular}{l|l}
\toprule
Pre-training task & MAE \\
\toprule
\emph{response}                                                 & $\textbf{50.65}\pm1.26$ \\
\emph{response} + \emph{elapsed\_time}                          & $54.86\pm1.64$ \\
\emph{response} + \emph{timeliness}                             & $52.91\pm1.38$ \\
\emph{response} + \emph{exp\_time}                              & $57.54\pm1.47$ \\
\emph{response} + \emph{inactive\_time}                         & $60.69\pm1.74$ \\
\emph{correctness}                                              & $51.36\pm0.97$ \\
\emph{correctness} + \emph{elapsed\_time}                       & $53.36\pm1.43$ \\
\emph{correctness} + \emph{timeliness}                          & $52.60\pm1.20$ \\
\emph{correctness} + \emph{exp\_time}                           & $54.36\pm1.62$ \\
\emph{correctness} + \emph{inactive\_time}                      & $55.04\pm1.58$ \\
\emph{response} + \emph{correctness}                            & $51.13\pm1.60$ \\
\emph{response} + \emph{correctness} + \emph{elapsed\_time}     & $52.15\pm1.43$ \\
\emph{response} + \emph{correctness} + \emph{timeliness}        & $53.05\pm1.81$ \\
\emph{response} + \emph{correctness}+ \emph{exp\_time}          & $53.09\pm1.25$ \\
\emph{response} + \emph{correctness} + \emph{inactive\_time}    & $56.41\pm1.72$ \\
\bottomrule
\end{tabular}
\label{tab:comparison_pre-training_tasks}
\end{table}

The best result was obtained under the pre-training task of predicting \emph{response} alone, which is slightly better than that of predicting \emph{correctness}, and both \emph{response} and \emph{correctness}.
Predicting correctness of student response is an important task in AIEd as can be seen from the large volume of studies about Knowledge Tracing.
Also, \cite{choi2020assessment} empirically showed that student response correctness is the most pedagogical interactive feature for academic test performance prediction.
However, rather than pre-training a model to predict whether a student correctly responded to a given exercise, the pre-training task of predicting student response itself injects more fine-grained information into the model, which leads to the more effective pre-training for academic test performance prediction.
Interestingly, the underperformed results were obtained when predicting \emph{elapsed\_time} or \emph{timeliness} in pre-training despite the benefits their information bring to several AIEd tasks \cite{feng2009addressing,zhang2017incorporating,shin2020saint+}.
We hypothesize that \emph{elapsed\_time} and \emph{timeliness} may introduce irrelevant noises and thus guide the model towards a direction inappropriate for academic test performance prediction.
In the case of \emph{exp\_time} and \emph{inactive\_time}, we observed that the generator failed to learn to predict their values when only given the interactive features listed in Section \ref{sec:transfer_learning}, which leads to unstable pre-training.
From these observations, in the following subsections, we conduct experimental studies based on the pre-training task of predicting \emph{response} alone.

\subsection{DPA vs. Baseline Methods}

\begin{table}[h]
\centering
\caption{Comparison of DPA with baseline methods.}
\begin{tabular}{l|l|l}
\toprule
Pre-training method & Fine-tuning model & MAE \\
\toprule
No pre-training     & MLP                   & $82.89\pm3.23$ \\
                    & BiLSTM                & $84.05\pm2.06$ \\
                    & Transformer encoder   & $107.06\pm2.52$ \\
                    & Performer encoder     & $81.76\pm1.24$ \\
\midrule
AE                  & MLP                   & $79.46\pm1.15$ \\
                    & BiLSTM                & $85.64\pm1.89$ \\
                    & Transformer encoder   & $75.13\pm3.10$ \\
                    & Performer encoder     & $64.80\pm1.43$ \\
\midrule
AM                  & MLP                   & $77.17\pm2.14$ \\
                    & BiLSTM                & $58.16\pm1.28$ \\
                    & Transformer encoder   & $57.16\pm2.08$ \\
                    & Performer encoder     & $52.79\pm1.39$ \\
\midrule
DPA                 & MLP                   & $77.24\pm1.59$ \\
                    & BiLSTM                & $57.59\pm1.76$ \\
                    & Transformer encoder   & $55.99\pm1.62$ \\
                    & Performer encoder     & $\textbf{50.65}\pm1.26$ \\

\bottomrule
\end{tabular}
\label{tab:comparison_baselines}
\end{table}

We compare DPA with the following pre-training methods:
\begin{itemize}
\item No pre-training: We train the fine-tuning models only on the fine-tuning dataset.
\item Autoencoding: Autoencoding (AE) is a generative pre-training method widely used across different domains of machine learning including AIEd \cite{guo2015predicting,ding2019transfer}.
Given an unmasked interaction sequence, AE pre-trains a model to reconstruct the input interaction sequence.
\item Assessment Modeling: Assessment Modeling (AM) \cite{choi2020assessment} is the previous state-of-the-art generative pre-training method for academic test performance prediction.
In AM, a model takes a masked interaction sequence as an input and is pre-trained to predict masked features.
AM is exactly the same as fine-tuning the pre-trained generator in DPA.
\end{itemize}
Also, we investigate whether DPA is effective with the following different fine-tuning models:
\begin{itemize}
\item MLP: Multi-Layer Perceptron (MLP) is stacks of simple fully-connected layers. 
Given an interaction sequence, interaction embedding vectors of all time-steps are summed together to compute a fixed-dimensional vector which is fed to a series of the fully-connected layers.
\item BiLSTM: Bi-directional Long Short-Term Memory (BiLSTM) is a model widely used for time series data prediction tasks.
The global max pooling layer is applied on top of the BiLSTM layer to obtain a fixed-dimensional intermediate representation from an input sequence of varying length.
\item Transformer Encoder: Transformer Encoder is a series of several identical layers composed of a multi-head self-attention layer with the softmax attention kernel and a point-wise feed-forward layer. 
We set the Transformer encoder’s attention window size to 512 due to the out of GPU memory occuring when training the Transformer encoder of 1024 attention window size on our single GPU machine.
\end{itemize}
As described in Table \ref{tab:comparison_baselines}, transferring the pre-trained knowledge brings better results in most cases, and the best result is obtained from DPA.
Especially, when the Performer encoder, the best performing fine-tuning model, is used as the fine-tuning model, DPA reduces MAE by 4.05\%, 21.84\%, and 38.05\% compared to AM, AE, and No pre-training, respectively.
Among the baseline pre-training methods excluding No pre-training, the worst result is obtained from AE beacuse the pre-training task of AE is much easier than that of AM and DPA.
We observed that the loss curve of AE converged to near zero within the first pre-training evaluation.

\subsection{Robustness to Increased Label-scarcity}

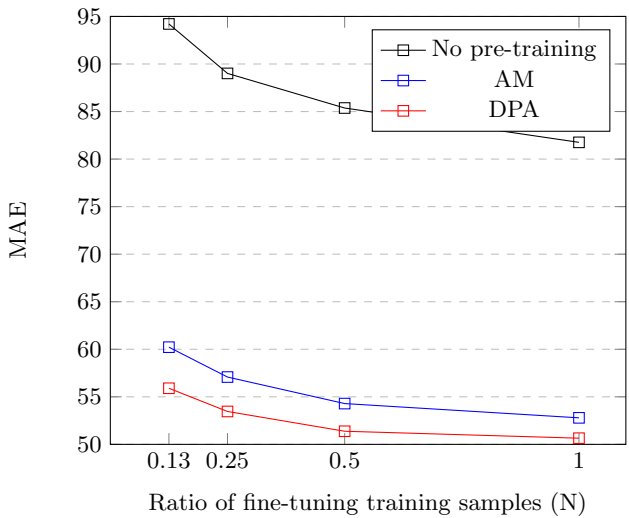
\begin{figure}[t]
\centering
\begin{tikzpicture}
\begin{axis}[
    xlabel={Ratio of fine-tuning training samples (N)},
    ylabel={MAE},
    xmin=0, xmax=1.1,
    ymin=50, ymax=95,
    xtick={1/8, 1/4, 1/2, 1},
    ytick={50, 55, 60, 65, 70, 75, 80, 85, 90, 95},
    legend pos=north east,
    ymajorgrids=true,
    grid style=dashed,
]
\addplot[
    color=black,
    mark=square,
    ]
    coordinates {
    (1/8, 94.21) (1/4, 89.01) (1/2, 85.37) (1, 81.76)
    };
    \addlegendentry{No pre-training}
\addplot[
    color=blue,
    mark=square,
    ]
    coordinates {
    (1/8, 60.22) (1/4, 57.08) (1/2, 54.29) (1, 52.79)
    };
    \addlegendentry{AM}
\addplot[
    color=red,
    mark=square,
    ]
    coordinates {
    (1/8, 55.90) (1/4, 53.46) (1/2, 51.38) (1, 50.65)
    };
    \addlegendentry{DPA}
\end{axis}
\end{tikzpicture}
\caption{The black, blue, and red lines represent MAEs for No pre-training, AM, and DPA, respectively, when the number of fine-tuning training samples becomes 1/2, 1/4, and 1/8 of the entire dataset.}
\label{fig:label_scarcity}
\end{figure}

Since the motivation behind our proposal of DPA is the label-scarcity problem, we investigate how MAE changes at varying degrees of label-scarcity.
Figure \ref{fig:label_scarcity} and Table \ref{tab:comparison_label_scarcity} describe the results when using 1/2, 1/4, and 1/8 of the total number of fine-tuning training samples.
In all degrees of label-scarcity, DPA consistently outperforms AM.
Also, DPA fine-tuned on 1/2, 1/4, and 1/8 of the dataset outperforms AM fine-tuned on the entire dataset, 1/2, and 1/4 of the dataset, respectively, which shows that DPA is more robust to label-scarcity than AM.
Compared with No pre-training, the gap between No pre-training and the other two pre-training methods increases as the number of labels becomes scarce.
Furthermore, the other two pre-training methods fine-tuned on 1/8 of the dataset outperform No pre-training fine-tuned on the entire dataset.

\begin{table}[t]
\centering
\caption{Comparison of DPA with AM and No pre-training at varying degrees of label-scarcity.}
\begin{tabular}{l|l|l|l}
\toprule
N & No pre-training & AM & DPA \\
\toprule
1/8     & $94.21\pm8.40$ & $60.22\pm1.86$ & $55.90\pm1.97$ \\
1/4     & $89.01\pm2.14$ & $57.08\pm1.75$ & $53.46\pm1.45$ \\
1/2     & $85.37\pm1.15$ & $54.29\pm1.50$ & $51.38\pm1.16$ \\
Full    & $81.76\pm1.24$ & $52.79\pm1.39$ & $50.65\pm1.26$ \\
\bottomrule
\end{tabular}
\label{tab:comparison_label_scarcity}
\end{table}

\subsection{Analysis of Predictions by Score Distribution}

\begin{figure}[t]
\centering
\includegraphics[width=0.5\textwidth]{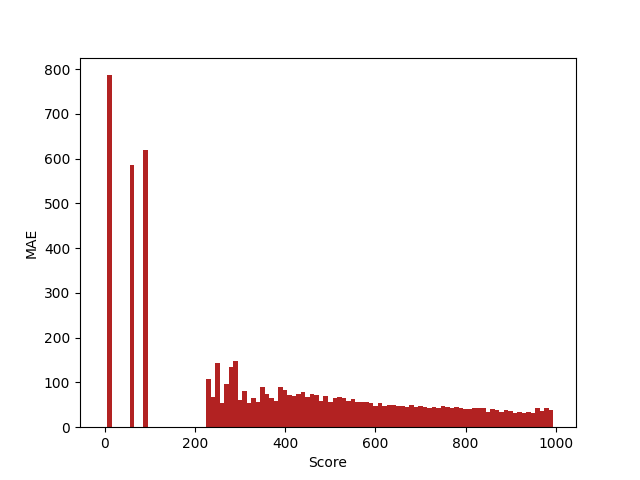}
\caption{MAEs by score distribution.}
\label{fig:score_mae}
\end{figure}

Figure \ref{fig:data_distribution} described that score labels are mainly distributed over the specific ranges.
We investigate how this biased distribution of score labels affects model predictions.
The results are described in Figure \ref{fig:score_mae}.
As expected, DPA severely underperforms when the score labels are below 200.
Although these are natural results from a machine learning perspective, this is a serious problem from a perspective of educational service because students whose scores are lower than 200 are inaccurately diagnosed their academic status.
It is also against the equity of education since not all students can receive the same level of educational service.
There may be various research directions to solve this problem, such as generating pseudo labels, measuring prediction uncertainties, or even collecting more score labels. 
However, we don’t go deeper into it any further and leave it as a future work.

\subsection{Analysis of DPA}

\begin{figure*}[ht]
\centering
\includegraphics[width=1\textwidth]{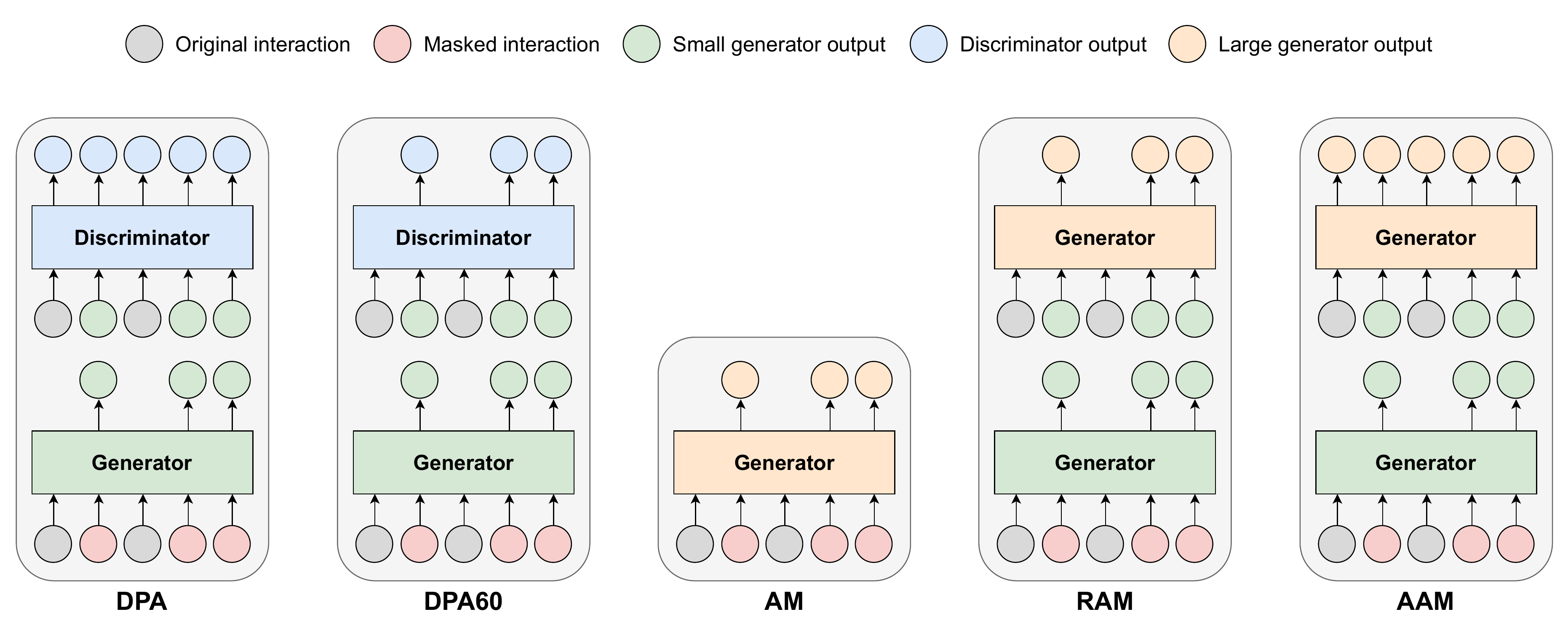}
\caption{Graphical descriptions of DPA, DPA60, AM, RAM, and AAM.}
\label{fig:ablation}
\end{figure*}

The previous experimental results showed that DPA makes better predictions and is more robust to label-scarcity than AM.
However, where the gains from DPA are coming from is not obvious.
We investigate what makes DPA outperform AM by comparing DPA and AM with the following set of ablation pre-training methods:
\begin{itemize}
\item DPA60\%: DPA60\% is the same as DPA except the discriminator loss tokens only come from the masked interactions.
Since we set the masked interaction ratio to 60\%, DPA60\% pre-trains the discriminator with 40\% fewer loss tokens than DPA.
\item RAM: Unlike DPA where there are the generator and the discriminator, in Replaced Assessment Modeling (RAM), there are two generators of different sizes, a small and large generator.
The small generator is pre-trained in the same way as AM.
The large generator takes a replaced interaction sequence which is generated by replacing the masked features with the small generator’s outputs as an input, and is pre-trained to predict the masked features.
After the pre-training, we throw away the small generator and fine-tune the large generator.
The sizes of the small and large generator are the same as those of DPA’s generator and discriminator, respectively.
\item AAM: All-tokens Assessment Modeling (AAM) is the same as RAM except the large generator is pre-trained to predict all features rather than just predicting the masked features.
\end{itemize}
Figure \ref{fig:ablation} depicts graphical description of each pre-training method.
The results are described in Table \ref{tab:comparison_ablation}.
In the following subsections, we analyze the results in aspects of pre-train/fine-tune discrepancy due to the mask token, discriminative vs. generative pre-training, and sample efficiency.

\begin{table}[t]
\centering
\caption{Comparison of DPA and AM with other ablation pre-training methods.}
\begin{tabular}{l|l}
\toprule
Pre-training method & MAE \\
\toprule
AM      & $52.79\pm1.39$ \\
RAM     & $52.64\pm1.21$ \\
DPA60   & $51.87\pm1.63$ \\
AAM     & $51.07\pm1.23$ \\
DPA     & $50.65\pm1.26$ \\
\bottomrule
\end{tabular}
\label{tab:comparison_ablation}
\end{table}

\subsubsection{Pre-train/Fine-tune Discrepancy Due to Mask Token}

During pre-training, the generator in AM sees the mask token which does not appear in fine-tuning, leading to pre-train/fine-tune discrepancy.
\cite{yang2019xlnet,clark2020electra} also raised the same issue found in the MLM pre-training method proposed in \cite{devlin2018bert}.
Since the discriminator in DPA does not see the mask token both in pre-training and fine-tuning, DPA does not suffer from the pre-train/fine-tune discrepancy, making it necessary to examine how much gain is obtainable from this issue.
Comparing RAM and AM, removing the pre-train/fine-tune discrepancy slightly reduces MAE by 0.28\%.

\subsubsection{Discriminative vs. Generative Pre-training}

The discriminator in DPA and the generator in AM are pre-trained with different objectives.
For each token to be predicted, the discriminator pre-training loss comes from a discrimination error between the discriminator’s output and an originality of the token.
On the other hand, the generator is pre-trained by generation error between the generator’s output and an identity of a token to be predicted.
Comparing DPA with AAM, and DPA60 with RAM, the discriminative pre-training objective reduces MAE by 0.82\% and 1.46\%, respectively, over the generative pre-training objective.

\subsubsection{Sample Efficiency}

The loss tokens for pre-training the discriminator in DPA come from all interactions in the input interaction sequence.
However, the generator in AM is pre-trained with loss tokens only coming from the masked interactions.
Considering the masked interaction ratio is 60\%, DPA is more sample efficient than AM since the discriminator is pre-trained with 40\% more loss tokens than the generator.
Comparing DPA with DPA60, and AAM with RAM, sample efficient pre-training brings reduction in MAE by 2.35\% and 2.89\%, respectively, showing that the sample efficiency contributed most to the improvement from AM to DPA.

\section{Conclusion}
In this paper, we proposed DPA, a transfer learning framework with discriminative pre-training tasks for academic performance prediction.
Our experimental results showed the effectiveness of DPA for the label-scarce academic performance prediction task over the previous state-of-the-art generative pre-training method.
Avenues of future research include investigating more effective pre-training tasks for academic performance prediction and pre-train/fine-tune relations in AIEd.

\bibliographystyle{abbrv}
\bibliography{sigproc} 

\balancecolumns

\end{document}